\documentclass[aps,prb,twocolumn,superscriptaddress,showpacs]{revtex4}
\usepackage{amssymb,amsmath}
\usepackage{graphicx}

\def\G{\Gamma}
\def\g{\gamma}
\def\O{\Omega}
\def\k{\bf k}
\newcommand{\dt}[1]{\Delta t_{1#1}}
\newcommand{\coh}{P_{01}}
\newcommand{\PL}[1]{P_{#1}^L}
\newcommand{\PLa}[1]{P_{#1}^{L\ast}}
\newcommand{\eq}[1]{Eq.~(\ref{#1})}

\newcommand{\fig}[1]{Fig.~\ref{#1}}

\begin{document} 

\title{Observation of  inter-Landau-level quantum coherence
in semiconductor quantum wells}

\author{K. M. Dani}

\affiliation{ 
Materials Science and Technology Division, MST-10, 
Los Alamos National Laboratory, Los Alamos, New Mexico 87545, USA}

\affiliation{Department of Physics, 
University of California at Berkeley,
Berkeley, California 94720, USA}

\affiliation{
Materials Sciences Division,
E.O. Lawrence Berkeley National Laboratory, 
Berkeley, California 94720, USA}

\author{I. A. Cotoros}

\affiliation{Department of Physics, 
University of California at Berkeley,
Berkeley, California 94720, USA}

\affiliation{
Materials Sciences Division,
E.O. Lawrence Berkeley National Laboratory, 
Berkeley, California 94720, USA}

\author{J. Wang} 

\affiliation{
Materials Sciences Division,
E.O. Lawrence Berkeley National Laboratory, 
Berkeley, California 94720, USA}

\author{J. Tignon}

\affiliation{Laboratoire Pierre Aigrain, 
Ecole Normale Sup\'erieure, F-75005 Paris, France.}

\author{D. S. Chemla}

\affiliation{Department of Physics, 
University of California at Berkeley,
Berkeley, California 94720, USA}

\affiliation{
Materials Sciences Division,
E.O. Lawrence Berkeley National Laboratory, 
Berkeley, California 94720}

\author{E. G. Kavousanaki}

\affiliation{
Chemistry Department, University of California, Irvine, 
California 92697, USA}

\affiliation{
Institute of Electronic Structure \& Laser, Foundation
for Research \& Technology-Hellas,
and Department of Physics, University of Crete, 
GR-71003 Heraklion, Greece}

\author{I. E. Perakis}

\affiliation{
Institute of Electronic Structure \& Laser, Foundation
for Research \& Technology-Hellas,
and Department of Physics, University of Crete, 
GR-71003 Heraklion, Greece}

\date{\today}

\begin{abstract}
Using three-pulse four--wave--mixing femtosecond
spectroscopy, we excite a non--radiative 
coherence between the discrete Landau levels 
of an undoped quantum well
and study its dynamics.
We observe quantum beats that reflect the time evolution 
of the coherence
between the two lowest Landau level magnetoexcitons. 
We interpret our observations using a 
 many-body theory
and find that the inter--Landau level coherence 
decays with a new time constant, 
substantially longer than 
the corresponding  interband magnetoexciton dephasing times. 
Our results indicate a new intraband
excitation dynamics that cannot be described in terms of 
 uncorrelated  interband 
excitations.

\end{abstract}


\maketitle

Quantum coherences between discrete states, 
formed by creating a superposition 
with well-defined relative phase, 
are central for manipulating matter on a quantum level and
can provide   the basis of schemes for  
information processing. 
Raman coherences in  atomic systems lead
to non--linear optical effects with potential technological importance,  
such as electromagnetically induced 
transparency and lasing without inversion. \cite{eit}
For applications, it is  desirable to observe and manipulate analogous
coherences in  semiconductors. Quantum beats 
due to coherence between heavy and light hole 
valence band  states, 
\cite{inter-val} as well as 
collective excitations in the quantum Hall system \cite{QHE}
and spin excitations in quantum dots, \cite{QD}
 have been reported.
Standard two--pulse four--wave--mixing (FWM) 
experiments do not access directly the 
Raman coherence, 
which can be inferred by simultaneously measuring the pump--probe 
signal
\cite{p-p} or by using 
three--pulse wave--mixing. 
\cite{cundiff,QHE}
Recently, Spivey {\em et. al.} 
\cite{cundiff} demonstrated faster dephasing of the 
coherence between heavy  and light hole excitons
than the corresponding interband dephasing.  

\begin{figure}[t]
\includegraphics[width=8.6cm]{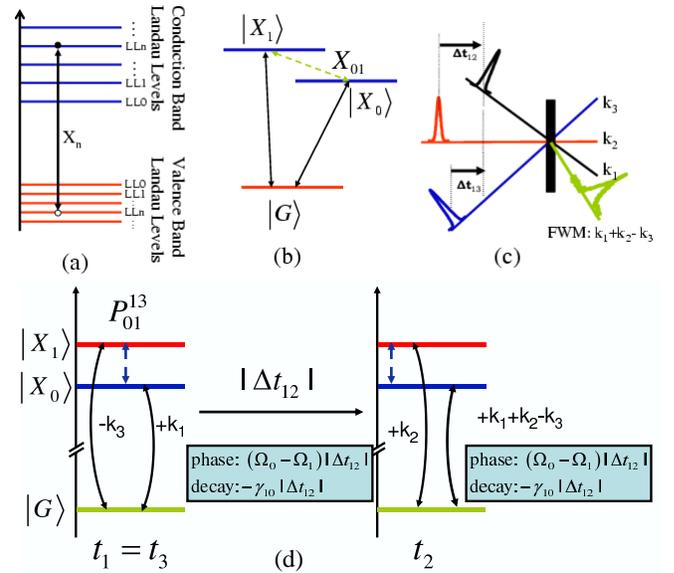}
\caption{(Color online)
Schematics of 
(a) X$_n$ states, 
(b)  X$_{01}$ Raman coherence,
(c)  three-pulse FWM, and
(d) nonlinear process describing the X$_{01}$ coherence 
contribution.
}   
\label{schem}
\end{figure}

\begin{figure*}
\includegraphics[width=15cm]{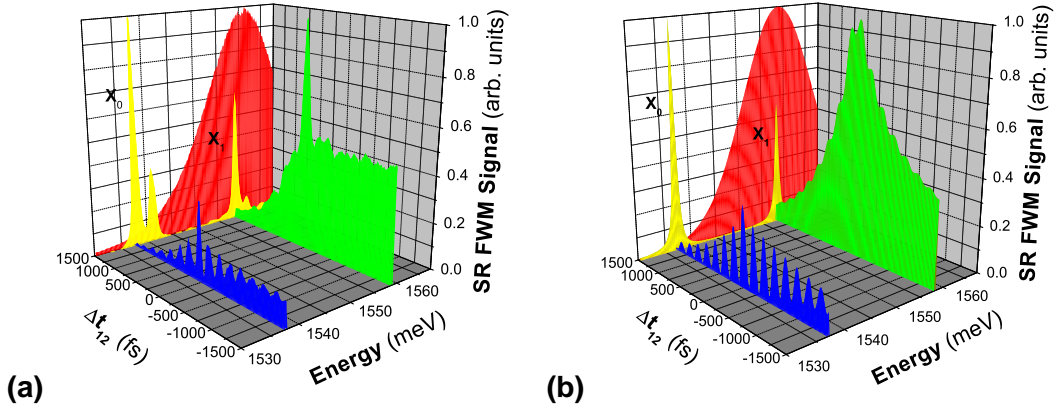}
\caption{ (Color online)
Three-pulse FWM signal along the $\dt{2}$ axis for large 
X$_1$ over X$_0$ excitation ratio:
(a) experiment, and (b) theory.
The experimental FWM signals from X$_n$ were obtained using a 1 nm
bandwidth interference filter and assigned to the respective
linear absorption peaks in this three--dimensional 3D representation. 
Backpanel: Linear absorption and optical pulse intensity.} 
\label{3d}
\end{figure*}

Here we create and study long--lived 
 coherence between magnetoexcitons
in semiconductor quantum wells (QWs), which we 
control with  a  perpendicular magnetic (B) field. 
The QW confinement discretizes the eigenstates along 
the z--axis (growth direction). The 
B--field results in quasi-confinement within the x--y plane, thus 
discretizing the eigenstates into Landau levels (LL). 
Continuum states are suppressed and the 
resulting discrete spectrum can be tuned with the B--field.   
Such  effective zero--dimensional confinement 
 opens new 
possibilities for creating and manipulating  coherence.
Long-lived intraband coherences  could
 be useful in future applications of quantum technology.
Also, understanding the coherence dynamics in 
QWs subject to a B--field 
is a necessary step towards a comprehensive picture of 
the quantum 
dynamics in the quantum Hall effect regime.\cite{QHE,perakis} 

In semiconductors, 
the Coulomb interaction 
leads  to effects such as exciton-exciton 
correlations, \cite{Ultra1,Ultra2,kner}
while exciton coupling to the environment 
gives both dephasing and new coherences.\cite{DCTS}   
In  
many--body systems, it is not easy to treat such complex 
correlations  
theoretically. \cite{perakis}
Thus, the  measurement of quantities that characterize the
coherence dynamics
gives 
valuable information on the non--equilibrium properties 
 of complex 
systems.

Here we investigate the dynamics of  interaction--induced 
inter--LL coherence in an undoped QW subject to a large 
B--field.
 We photoexcite X$_0$ and X$_1$ magnetoexcitons (X$_n$ 
consists of an electron in the $n$-th conduction-band LL
and a hole in the $n$-th valence-band LL, \fig{schem}a) and create  
a X$_0 \leftrightarrow$X$_1$  coherence, X$_{01}$
(\fig{schem}b). We identify a three-pulse 
FWM signal  that reflects the dynamics 
of X$_{01}$  and displays quantum beats with 
a new decay time. 
Using a  
many-body theory, \cite{perakis}
we identify the source of these beats and 
extract the dephasing rate of  X$_{01}$.
We find that, unlike for 
 uncorrelated magnetoexcitons,  this rate is
 substantially smaller than 
the  sum of the magnetoexciton dephasing rates.

\begin{figure}[b]
\includegraphics[width=8.4cm]{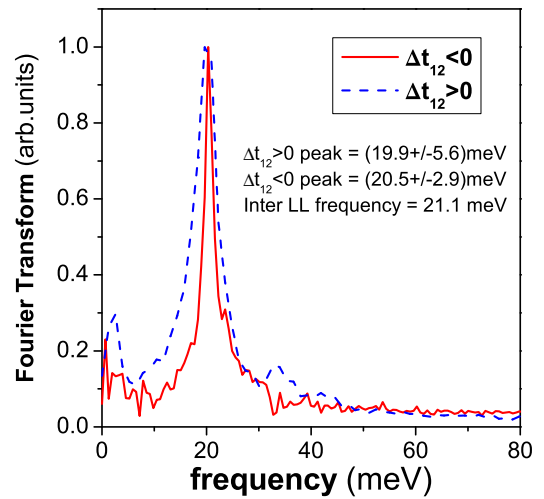}
\caption{ (Color online)
Fourier transform of the oscillations for $\dt{2}<0$ (solid line)
and $\dt{2}>0$ (dashed line). Both show a single peak
at $\O_1-\O_0$, 
but the  $\dt{2}<0$
linewidth (2.9meV) is significantly smaller that the
$\dt{2}>0$  one (5.6meV).
}   
\label{depend}
\end{figure}

\begin{figure*}
\includegraphics[width=15cm]{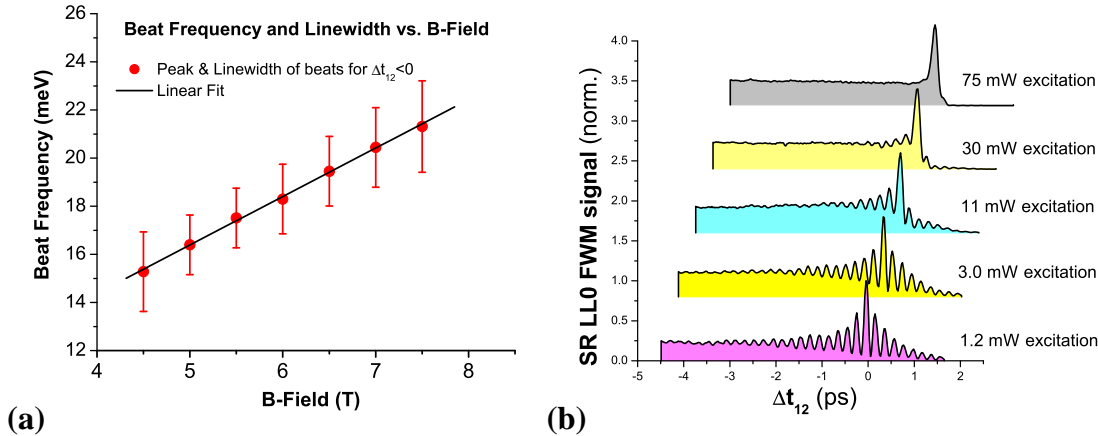}
\caption{ (Color online)
(a) B--field dependence of the oscillation frequency 
along the $\dt{2}$ axis. 
Error bars: 
linewidths in the oscillation Fourier transforms for  $\dt{2}<$0
(see Fig.\ref{depend}).
(b) Intensity dependence of the FWM signal along the $\dt{2}$ axis. 
The curves were normalized and shifted for clarity.
The $x$-offset is 0.3 ps, and the $y$-offset is 0.8.
}   
\label{intensity}
\end{figure*}

We study a ten well, undoped GaAs QW
structure, with 14 nm thick GaAs layers sandwiched between 10 nm
thick layers of Al$_{0.3}$Ga$_{0.7}$As. The sample is kept at 1.5-4$^\circ$K 
in a split-coil magneto-optical cryostat. A B--field 
(B=0-7 T) is applied along the QW  growth direction.
We excite the sample with three $100$fs pulses of 
right-circularly polarized ($\sigma_+$) light (\fig{schem}c).
 These 
pulses propagate along
directions $\k_{1}$, $\k_{2}$, and 
$\k_{3}$, with a time delay $\dt{2}$ ($\dt{3}$) between pulse $\k_{1}$ and 
$\k_{2}$ ($\k_{3}$). For negative values of the above delays, pulse $\k_{1}$ 
arrives first. We study the transient signal in the 
background free direction $\k_{1}+\k_{2}-\k_{3}$. Using an interference 
filter, we spectrally resolve this signal and separate the
X$_0$ and X$_1$ responses. We measure the signal intensity 
from each state (spectrally resolved FWM) as function of 
$\dt{2}$ and $\dt{3}$. As explained below, 
the $\dt{2}$ axis ($\dt{3}=0$)
accesses the dynamics of the intra--band coherence, while the 
$\dt{3}$  axis measures the interband polarization dephasing. 

\fig{schem}d shows a schematic of the 
 FWM signal due to the X$_{01}$
coherence along the negative $\dt{2}$ axis.
To contribute in the $\k_{1}+\k_{2}-\k_{3}$ 
direction, the inter--LL excitation X$_{01}$ must be created by 
either  $\k_{1}$ and $\k_{3}$ 
or $\k_{2}$ and $\k_{3}$ pulses.
In \fig{schem}d, pulses $\k_{1}$ and $\k_{3}$ arrive together 
($\dt{3}=0$), and create the X$_{01}$ coherence, which evolves for a 
time $|\dt{2}|$ before it is probed by pulse $\k_{2}$. 
During this $|\dt{2}|$ time interval, the coherence accumulates 
a phase at frequency 
$\O_{0}-\O_{1}$ and decays with a rate $\g_{01}$, both 
reflected in the FWM dependence on $\dt{2}$
 ($\O_{n}$ is the energy
of X$_n$). 
Thus, for $\dt{2}<0$,
we can access the X$_{01}$ coherence  dynamics, 
while, for $\dt{2}>0$, pulse $\k_{2}$ 
arrives first and the FWM signal  reflects the 
dynamics of the interband polarization created by $\k_{2}$.

\fig{3d}a shows the FWM intensity from both X$_0$ and X$_1$
along the 
$\dt{2}$ axis. We largely excite 
X$_1$ over X$_0$ (see backpanel of \fig{3d}) 
in order to suppress the Pauli blocking [phase 
space filling (PSF)] contribution at X$_0$.
We then see  a very small X$_0$ 
signal (as compared to X$_1$) with striking beats. As discussed above,
the 
negative and positive $\dt{2}$ axes reflect different dynamics, 
so we analyze the decay rates separately. We 
subtract a constant (exponential) background from the negative 
(positive) axis and take the Fourier transform of the resulting signal 
(\fig{depend}). In both cases, we see a strong peak at energy 
$\O_{1}-\O_{0}$, where $\O_1$ and $\O_0$ are obtained
from the linear absorption spectrum (back panel of \fig{3d}a). 
However, we see a large 
difference in the linewidths obtained by
fitting a Lorentzian to the peaks: 2.9 meV for the negative 
side vs. 5.6 meV for the positive. Therefore, the 
beats decay  slower for $\dt{2}<0$. This asymmetric decay 
allows us to identify the $X_{01}$ decoherence time.

\fig{intensity}a shows the beat frequencies 
and linewidths extracted from the X$_0$ FWM signal
along the negative $\dt{2}$ axis
for various B--fields. 
 We see a linear dependence of the beat
frequency on B, as expected 
for  large B--fields (where the cyclotron  energy 
exceeds the Coulomb energy \cite{x-x,shah}) 
from the inter--LL
energy $\O_{1}-\O_{0}$.
From the slope of \fig{intensity}a we extract  
an e--h reduced mass of 0.058$\pm$0.001$m_e$, 
which, for electron mass 
of 0.066$m_e$, 
corresponds to the heavy--hole mass of 0.498$m_e$.  
 On the other hand, we do not observe any 
substantial linewidth changes with B. 
We also studied the changes in the X$_0$ FWM signal for increasing 
photoexcitation intensity (\fig{intensity}b). The 
decoherence times decrease as the photoexcited carrier density increases, 
while the asymmetric beat decay and overall temporal profile remain the same.

We analyze our results using a 
 many-body theory \cite{perakis} based on 
the dynamics controlled truncation scheme (DCTS). \cite{DCTS}
We expand in terms of the 
optical field in order to decrease the number of independent 
dynamical variables  and separate out
the correlated contributions to the third--order non--linear 
optical response.
We consider $\sigma_+$ 
optical pulses and 
include for simplicity only the photoexcited X$_0$ and X$_1$ 
states. We use the standard  Hamiltonian 
that treats  the Coulomb interactions between  
carriers in a B--field.
\cite{perakis}
For $\sigma_+$ polarized light, the only dipole--allowed optical 
transition is from the ($j=3/2$, $m_j=-3/2$) valence band into the 
(1/2,-1/2) conduction band.\cite{fromer} 
This, as well as the measured linear dependence of the 
beat frequency on the B--field,
allows us to consider a
simple two band semiconductor model and assume the heavy hole 
mass of 0.498$m_e$.

The FWM signal from X$_n$ is described by the
polarization 
$P_n=\langle X_n\rangle$. We derive the following equation of motion for 
$P_0$
(ignoring the non-resonant terms):
\begin{eqnarray}\label{p0}
i\partial_t P_0 &=& (\O_{0}-i\G_{0}) P_0
+2 \mu E(t)(\PL{0} \PLa{0} + N_0)\nonumber\\
&&-2V_{01} \PL{0} \left(\PLa{1}\PL{1} + N_1 \right) 
-2V_{01} \PL{1}\coh 
\end{eqnarray}
where $\O_{0}$ and $\G_{0}$ are the energy and  dephasing
rate of  X$_0$, $\mu$ the interband transition matrix element,
$E(t)$ the optical pulse, $V_{01}$ the 
X$_0$--X$_1$ interaction,\cite{perakis} $P_n^L$ the linear X$_n$ polarization, 
$\coh=\langle X_{01}\rangle=
\langle |X_1\rangle\langle X_0|\rangle-\PL{0}\PLa{1}$
the X$_{01}$ Raman coherence, 
and $N_n$ the incoherent X$_n$ density. The second term on the right hand side 
(rhs)
of \eq{p0} is due to PSF. 
The last two terms,  
due to the Coulomb 
interaction $V_{01}$,
give a nonlinear coupling 
of X$_0$ and 
 X$_1$.
Setting $N_n=\coh=0$ recovers the semiconductor Bloch equations 
in  a magnetic field.\cite{x-x}
We ignored 
the bi--magnetoexciton correlations \cite{shah,perakis}
since our FWM
 signal along the  negative $\Delta t_{13}$ axis, 
generated by these correlations, \cite{kner,shah}  
is suppressed and decays fast.

\begin{figure}[t]
\includegraphics[width=8.4cm]{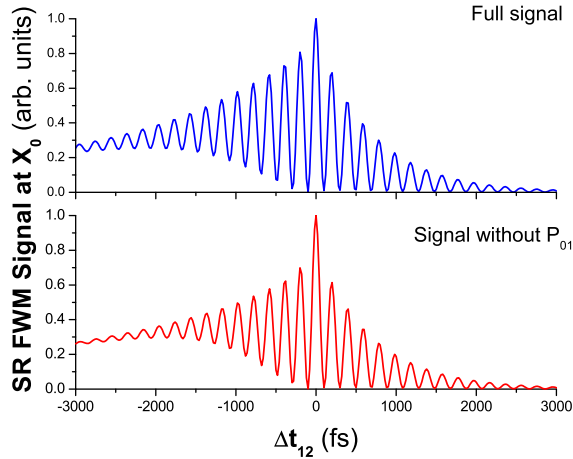}
\caption{ (Color online)
Numerical simulation of the X$_0$ FWM signal 
with and without the contribution of the X$_{01}$ coherence.
}
\label{terms}
\end{figure}

The DCTS showed that, 
in the absence of 
correlations mediated by a bath, 
 the dynamics can be described in terms of 
interband amplitudes only. \cite{DCTS} 
It is then impossible 
to obtain lifetimes 
longer than the magnetoexciton dephasing times. 
Here the key  variable 
is 
$\coh$,\cite{DCTS,perakis}
\begin{equation}
\label{coh}
i\partial_t \coh=(\O_{0}-\O_{1}-i\g_{01})\coh + 
i(\G_{0}+\G_{1}-\g_{01})\PL{0} \PLa{1},  
\end{equation}
which evolves 
with frequency $\O_{0}-\O_{1}$ and decays with  rate $\g_{01}$
(as in the $|\dt{2}|$ interval of  \fig{schem}d). The second
(source) term
is due to the coupling 
to a bath, 
characterized
by dephasing rates.\cite{DCTS}
It arises 
when 
the  X$_{01}$ dephasing rate, $\g_{01}$,
deviates from the sum of 
the X$_{n}$ dephasing rates, $\Gamma_0+\Gamma_1$.
In the case of uncorrelated
interband transitions,
the inter--LL coherence is given by the 
magnetoexciton amplitude 
product $\PL{0} \PLa{1}$, which decays with a rate 
$\Gamma_0+\Gamma_1$. This Hartree--Fock result 
gives the  
third term on the rhs of  
Eq.(\ref{p0}).\cite{x-x}
$\g_{01}\ne\Gamma_0+\Gamma_1$ implies
correlations, mediated by the bath, 
between the interband transitions.
The dynamics of intraband and interband 
coherences can then be distinguished. 

A microscopic calculation of $\g_{01}$ and $\Gamma_n$ requires 
 complete knowledge of the coupling to the bath. Ref.[\onlinecite{perakis}] 
analyzed
the Coulomb coupling to a
cold electron gas. 
Ref.[\onlinecite{DCTS}] 
demonstrated  different time evolution of inter--band and intra--band  
variables  due to exciton--phonon dynamics.
In our system, dephasing arises from the interplay 
between phonon--carrier and  carrier--carrier scattering and 
the disorder, which breaks the LL degeneracy, 
leading to a finite 
LL width. 
A complete theory of this interplay 
is lacking at present.
However, our results 
clearly show different time evolution of X$_n$ and X$_{01}$
due to the bath.

The numerical solution of 
our full equations, including 
nonresonant contributions,  
gives the FWM signal 
of \fig{3d}b. This
reproduces  the  experimental
features. 
The  signal due to the 
term $\propto P_1^{L} \coh$ in Eq.(\ref{p0}) 
reflects the phase accumulated
by  $\coh$ during  $|\dt{2}|$, while the term 
 $\propto P_0^{L}N_1$ gives 
$P_0\propto e^{\g_{D}\dt{2}}$,
where $\g_{D}$ is the  relaxation rate of the 
incoherent X$_1$ population.
We attribute our long--lived coherence to the beating 
of the two above 
contributions, with frequency  $\O_{1}-\O_{0}$, 
which decays at a rate of $\g_{01}+\g_{D}\sim\g_{01}$.
All other FWM contributions lead to 
oscillations that decay as
$\G_{0}+\G_{1}$ or faster. On the other hand, for positive 
$\dt{2}$, all beatings 
have frequency $\O_{1}-\O_{0}$ and decay as $\G_{0}+\G_{1}$.
\cite{explanation}
This is illustrated 
in \fig{terms}: without the $\coh$ contribution,
the oscillations for both positive and negative $\dt{2}$ decay
with $\G_{0}+\G_{1}$; however,  including
$\coh$ with $\g_{01}<\G_{0}+\G_{1}$,
the oscillations decay more slowly on the negative axis. 
With increasing photoexcited carrier density, 
carrier--carrier scattering
enhances $\gamma_{01}$ and $\Gamma_n$, so 
the quantum beats decay faster as the intensity increases (\fig{intensity}b).
However, the overall FWM temporal profile remains 
unchanged, reflecting 
 population relaxation 
with very small $\gamma_D$.
We extract from the $\dt{2}<0$ oscillation decay
an inter--LL
coherence dephasing rate of
$\g_{01}=2.9$ meV. 
Our work 
demonstrates tunable quantum dynamics between 
Coulomb--coupled discrete Landau levels.

This work was supported by the Office of Basic 
Energy Sciences of the US Department of Energy and by the EU 
STREP program HYSWITCH.


\begin{references}
\bibitem{explanation}
When $\dt{2}>0$ (and $\dt{3}=0$), pulse $\k_2$ arrives first. 
It creates a polarization $\PL{n}$ that 
evolves for time $t=\dt{2}$ with frequency $\O_n$ and decay rate
$\G_n$, until pulses $\k_1$ and $\k_3$ come in. Different processes
can then take place, but the FWM signal emitted in the $\k_1+\k_2-\k_3$
direction will only depend on $\dt{2}$ as $e^{-i\O_n\dt{2}}e^{-\G_n\dt{2}}$.
Thus, the only oscillations that can be created beat with frequency
$\O_1-\O_0$ and decay as $\G_0+\G_1$.
\end{references}
\end{document}